\documentclass[10pt,journal]{IEEEtran}
\usepackage[T1]{fontenc}
\usepackage{multicol}
\usepackage{amsmath}
\usepackage{amsfonts}
\usepackage{indentfirst}
\usepackage[dvips]{graphicx}
\usepackage{xspace}
\usepackage{epsfig}
\usepackage{times}
\usepackage{color}
\usepackage{mathrsfs,multirow}
\usepackage{ae}
\usepackage{amssymb}
\usepackage{url}
\usepackage{theorem}
\usepackage{geometry}
\usepackage{verbatim}
\makeatletter

\usepackage{latexsym, amssymb, amsmath, tabularx, epsfig, xspace}
\usepackage{psfrag}

\newcommand{\F}{\ensuremath{\mathbb{F}}}

\theoremstyle{plain}%[chapter]

\newtheorem{theo}{Theoreme}[section]

\makeatother
\begin{document}
\title{ A new zero-knowledge code based identification  scheme with reduced communication}
\author{Carlos Aguilar, Philippe Gaborit, Julien Schrek \\
%\author{}
%\institute{}
Universit\'e de Limoges, France.\\
%123, Av. Albert Thomas 87060 Limoges Cedex France\\
\{carlos.aguilar,philippe.gaborit,julien.schrek\}@xlim.fr
}

\thispagestyle{empty}
\maketitle
\thispagestyle{empty}

%\begin{center}
%  \today
%\end{center}

\begin{abstract}

%%%%%%%%%%%%%%%%%%%%%%%%%%%%%%
In this paper we present a new 5-pass identification scheme with asymptotic cheating probability $\frac{1}{2}$ based on the syndrome decoding problem. Our protocol is related to the Stern identification scheme but has a reduced communication cost compared to previous code-based zero-knowledge schemes, moreover our scheme permits to obtain a very low size of public key and secret key. 
The contribution of this paper is twofold, first we propose a variation on the Stern authentication scheme which permits to decrease asymptotically the cheating probability to 1/2 rather than 2/3 (and very close to 1/2 in practice) but with less communication. Our solution is based on deriving new challenges from the secret key through cyclic shifts of the initial public key syndrome; a new proof of soundness for this case is given
Secondly we propose a new way to deal with hashed commitments in zero-knowledge schemes based on Stern's scheme, so that in terms of communication, on the average, only one hash value is sent rather than two or three. 
Overall our new scheme has the good features of having a zero-knowledge security proof based on well known hard problem of coding theory, a small size of secret and public key (a few hundred bits),  a small  calculation complexity,  for an overall  communication cost of 19kb for authentication (for a $2^{16}$ security) and  a signature of size of 93kb (11.5kB)
(for security $2^{80}$), an improvement of $40\%$ compared to previous schemes based on coding theory.

%%%%%%%%%%%%%%%%%%%%%%%%%%%%%%%

\medskip

\end{abstract}

\section{Introduction}

The use of coding theory for public key cryptography was initiated by McEliece more than 30 
years ago, although the system has often be considered as too costly and impractical because
of the size of the public key, code-based cryptography has received much more attention
in recent years. Besides the fact that code-based cryptography can possibly resist
to a quantum computer, code-based systems have also inherent interests: they are very fast 
and are usually easy to implement compared to number theory based systems. 
Such features make code-based systems good candidates for low-cost cryptography.

There are two main types of code-based cryptosystems: systems with hidden structure
like the McEliece cryptosystems (analogous to RSA) and systems with no hidden
structure (analogous to discrete log -based cryptosystems) like for instance
the Stern code-based authentication scheme (\cite{Ste93}). 
This second type of system is not vulnerable
to structural attacks which are the main cause of attacks on McEliece-like cryptosystems.
In practice as for the Stern scheme, they have not been attacked beneath the usual
improvement on the attack of the underlying hard problem.

In the case of coding theory the underlying hard problem (the Syndrome decoding problem SD)
is now well studied and considered as very secure.

\smallskip

Code-based Zero-knowledge authentication schemes are very interesting since 
their security is directly related to a hard problem, moreover they can be turned into
signature schemes through the Fiat-Shamir paradigm.
Meanwhile there are two strong drawbacks for these schemes.
The first drawback is the size of the public key which can attain 
several hundred thousand bits and the second drawback is the size of the communication
induced by the cheating probability, more than 150kb in practice for a $2^{80}$ security level.

The first drawback was resolved in part by Gaborit and Girault \cite{GG} who proposed
to use structured matrices like double-circulant matrices (matrices of the form $(I A)$
for $A$ a random circulant matrix) to reduce the size of the public key
to only a few hundred bits. The second drawback, the high cost of communications, largely remains.
\medskip
%{\bf Contribution of the paper}
%
In this paper we make a step further to obtain a small communication cost,
 our new algorithm, 
with the same type of security than previous algorithm and small size of keys, permits 
to reduce the size of communications by 40\%. We propose two different improvements,
a first improvement relies on using the double-circulant structure to increase the number of possible challenges, and the generic second improvement consists in a better use of commitment by compressing
them. In practice it is now possible to sign for
a security level of $2^{80}$ with a signature of size 93kb rather than 155kb,
and to get identified for a security level of $2^{-16}$ with 20kb rather than 31kb.

\section{Background on code-based authentication schemes}

\subsection{Previous work}

There are severals protocols based on the syndrome decoding problem, we quickly survey the main advances in this area
The first efficient protocol was proposed by Stern \cite{Ste93}: 
his idea was a new way to prove the knowledge of a word with small weight and fixed syndrome. The idea consist of 
revealing one of the three statements, the adequate weight with a masked syndrome, the adequate syndrome with a wrong weight
or a way the weight and the syndrome can be masked. The $3$ challenges structure implies a cheating probability equal to 2/3 
instead of 1/2 for the well known scheme of Fiat-Shamir. The Stern protocol is also uncommon by the use 
of hash functions. In \cite{Ste93} Stern presents another protocol which aims at reducing the cheating probability to 1/2 by cutting 
the challenge step into 2 parts. Indeed, adding this challenge in the scheme prevents the prover to reveal the third statement 
and reduce the probability close to 1/2. The next improvement was a reduction of communication due to Véron in \cite{V95}, the 
reduction is due to a different formulation of the secret, which decreases the cost of communication
 but increases the size of the key. 
In \cite{GG}, Gaborit-Girault proposed to use particular compact matrices (doubly circulant matrices) in order
to obtain a very short public matrix.  The last improvement appeared with the protocol of Cayrel-Véron-El Yousfi
where the aim was to reduce the cheating 
probability to 1/2 as well as in the second protocol of Stern but using fields with cardinality higher than $2$.
Our protocol uses the V\'eron variation that we recall here.

\subsection{Scheme of Veron}

\begin{itemize}
\item[] {\bf private key} : $(e,m)$ with $e$ of weight $w$ and of length $n$ and $m$ a random element of $\F_2^k$.% of length $k$.
\item[] {\bf public key} : $(G,x,w))$ with $G$ a random matrix of size $k\times n$ and $x=e+mG$.
\end{itemize}

\begin{figure}[!h]\label{fig:Veron}
\centering
\fbox{
  \begin{minipage}{7.4cm}
    \begin{enumerate}
    \item ~[Commitment Step] $P$ randomly chooses $u\in \F_2^{k}$ and a permutation $\sigma$ of
 $\{1,2,\ldots,n\}.$
      Then $P$ sends to $V$ the commitments $c_{1}$, $c_{2}$ and $c_3$ such that :
      $$c_{1}=h(\sigma);\ c_{2}=h(\sigma((u+m)G));$$
$$c_{3}=h(\sigma(uG+x));$$
    \item ~[Challenge Step] $V$ sends $b \in \{0,1,2\}$ to $P.$
    \item ~[Answer Step] Three possibilities :
      \begin{itemize}
      \item if $b=0:$ $P$ reveals $(u+m)$ and $\sigma.$
      \item if $b=1:$ $P$ reveals $\sigma((u+m)G)$ and $\sigma(e).$
      \item if $b=2:$ $P$ reveals $u$ and $\sigma.$
      \end{itemize}
    \item ~[Verification Step] Three possibilities :
      \begin{itemize}
      \item if $b=0:$ $V$ verifies that $c_{1},c_{2}$ have been honestly computed.
      \item if $b=1:$ $V$ verifies that $c_{2},c_{3}$ have been honestly computed, and $wt(\sigma(e))=w$.
      \item if $b=2:$ $V$ verifies that $c_{1},c_{3}$ have been honestly computed.
      \end{itemize}
    %\item Iterate the steps $1$,$2$,$3$,$4$ until the expected security level is reached.
    \end{enumerate}
  \end{minipage}
} \caption{Protocol of Veron}
\end{figure}

\section{A new scheme}

We now give more details and a high level overview on our two improvements.

\smallskip

\subsection{High level overview: Increasing the number of challenges}

 At the difference of the Fiat-Shamir scheme in which the cheating probability is 1/2,
this probability is 2/3 for the Stern protocol. It comes from the fact that proving
that a prover knows a codeword of small weight with a given syndrome, means proving two facts:
the fact that the syndrome of the secret is valid and the fact that the secret
has indeed a small weight. This situation induces that if one adds a random commitment
there are always two possibilities for cheating among the three cases, notably
since the attacker knows the syndrome of the secret.

The small weight of the secret is proved by using a permutation and a bitwise XOR 
which permit to retrieve the syndrome thanks to the linearity of both operations. 
In all schemes based on syndrome decoding there is a statement of the form : 
$$\sigma(e) + v$$
Here $e$ is the secret of low weight, $\sigma$ a permutation and $v$ a mask. 
In the V\'eron scheme $v$ is equal to $\sigma((u+m)G)$ which is a good mask for $\sigma(e)$ with $u$ a random word   
and $v$ is a random word in the Stern scheme.   
The idea described in the scheme of Stern $5$ pass \cite{Ste93} and \cite{CVY} is that a 
variation of $e$ can prevent a dependence on $v$ and $\sigma$. So there is 
no need to test the construction of $v$ and $\sigma$ at the same time any more. The cheating probability 
is now close to 1/2, indeed there is now only two challenges possible for the second query. 

The variation on $e$ can be done in different ways, Stern used $e$ as a codeword of a 
Reed-Muller code, Cayrel et al. used a scalar multiplication, in our case we use a rotation of the 
two parts of $e$. Using this rotation we can deduce the syndrome of each permuted word 
thanks to the propriety of double circulant codes presented here
Let $H=[I|A]$, for $A$ a circulant matrix of length $k$ and let the syndrome $s=H.y^t$ for $y=(y_1,y_2)$
For $r$ a cyclic shift on $n$ positions we obtain: 
$$ s=H\cdot(y_1,y_2)^t \Leftrightarrow r(s)=H\cdot(r(y_1),r(y_2))^t.$$

Our construction therefore leads to $2k$ possible challenges: $k$ coming from the choice of the shift
and $2$ possibilities for the second query (compared to $3$ in the classical case)
An attacker can easily cheats for k challenges among the 2k possible, and we show
that it is not possible for an attacker to cheat for more than $k+i$ challenges (for $i$
a security parameter) without knowing the secret.

This cyclic permutation point of view is an efficient way to reduce the cheating probability close to 1/2 in 
a binary scheme and without rising the communication cost like it was done in the scheme 
of Stern $5$ pass or considering non binary alphabet like in Cayrel et al. which also leads to less interesting communications

\subsection{High level overview: Commitments compression}

In Stern's scheme (or V\'eron's scheme), the prover has first to  send $3$ commitments
composed of $3$ hash of different values: $c_1,c_2$ and $c_3$ in V\'eron's protocol (for instance). 
The sending of these three hashes comes
at a certain cost. Meanwhile one can remark that if the protocol works well, the Verifier
retrieves $2$ hash values among the $3$ hash values sent. This remarks shows that in fact
it possible to optimize the manipulation of these commitments. The Prover first needs
 to compute the three hash values as usual, but then rather than sending
the three hash values, he sends a hash of the three hash values.
After receiving the challenge of the Verifier the Prover knows that the Verifier
is able to recover 2 of the 3 hash values, then he answers to the challenge
as usual, but also adds to his answer the missing hash value.

In the verification step, if all worked correctly the Verifier is able to recover the
first commitment (the hash of the concatenation of the three hash values $c_1,c_2$ and $c_3$)
through the two hashed values he retrieved and the third one in the answer of
the Verifier. Overall only 2 hash values are sent rather than 3.

This idea can be generalized to the case
of sequenced rounds, in that case for each round the Prover sends only the missing hash
value when the two others are recovered by the Verifier. In that case only a general
commitment for all the rounds needs to be sent: a hash value of the sequence of all hash values
of the different rounds. This point of view is very efficient in particular for
signature for which the average number of hash values sent per round drops from  3 to 1

Moreover this way of proceeding in secure in the random oracle model, since an error
in the final hash value implies an error in one of the hash of the round sequence

\subsection{Description of the protocol}

We use the same notations and the same keys as in the scheme of Veron. 
\begin{itemize}
\item[] {\bf private key} : $(e,m)$ with $e$ of weight $w$ and of length $n$ and $m$ a random element of $\F_2^k$.% of length $k$.
\item[] {\bf public key} : $(G,x,w))$ with $G$ a random matrix of size $k\times n$ and $x=e+mG$.
\end{itemize}
For simplicity matter we describe the protocol in figure \ref{np2} only for the first improvement since the second one is generic.

\begin{figure}[!h]
\centering

\fbox{

  \begin{minipage}{7.4cm}
    \begin{enumerate}
    \item ~[First commitment Step] $P$ randomly chooses $u\in \F^{k}$ and a permutation $\sigma$ of 
 $\{1,2,\ldots,n\}$.
      Then $P$ sends to $V$ the commitments $c_{1}$ and $c_{2}$ such that : 
      $$c_{1}=h(\sigma);\ c_{2}=h(\sigma(uG));$$
    \item ~[First part of the challenge]  $V$ sends a value $0 \le r \le k-1$ (number of shifted positions) to $P$.
\item ~[Final commitment Step] $P$ build $e_r=Rot_r(e)$ and sends the last part of the commitment :
$$ c_{3}=h(\sigma(uG+e_r))$$
    \item ~[Challenge Step] $V$ sends $b \in \{0,1\}$ to $P.$
    \item ~[Answer Step] Two possibilities : 
      \begin{itemize}
      \item if $b=0:$ $P$ reveals $(u+m_r)$ and $\sigma.$
      \item if $b=1:$ $P$ reveals $\sigma( uG)$ and $\sigma(e_r)$ where $e_r=Rot_r(e)$.
      \end{itemize}
    \item ~[Verification Step] Two possibilities :
      \begin{itemize}
      \item if $b=0:$ $V$ verifies that $c_{1},c_{3}$ have been honestly computed.
    \item if $b=1:$ $V$ verifies that $c_{2},c_{3}$ have been honestly computed.
    and that the weight of $\sigma(e_r)$ is $w$. 
      \end{itemize}
    \end{enumerate}
  \end{minipage}
} \caption{New double-circulant protocol}
\label{np2}
\end{figure}

The verification protocol consists in a reconstruction of the hash value committed to 
the first step of the algorithm. 
In the first case, the first and the third hash values 
can be constructed and in the second case it concerns the second and the third hash values.  
The construction of hash value are obvious except $c_3$ in the $b=0$ case using the two answers, the word $u$ and the permutation $\sigma$. 
We just have to see that $c_3=\sigma(uG+x_r)$, with $x$ the public key shifted $r$ times.

\section{Security}\label{security}

In this section we first prove the ZK security of our scheme by using the usual zero-knowledge arguments and we also consider practical security.

\subsection{Completeness}

	The completeness is clear at the moment that we notice that the sending of the prover permit to generate the corresponding hash value. It's pretty clear when wee see the verification scheme. 

\subsection{Soundness}

We prove here that a malicious prover cannot be authenticated with probability much higher than $\frac{1}{2}$. We introduce a new parameter $i$ to compute
a trade-off between the cheating probability, security cost and communication cost.
The idea of the proof is to prove that someone who can anticipate more than $k+i$ challenges can also retrieve the secret key with a good probability, depending on $i$. We use
the verification algorithm of the protocol to obtain necessarily conditions for cheating. The end of the proof consists in choosing a high enough parameter $i$ such that, with a good probability the only solution with a good condition is the secret key.

\begin{theo}

	If a prover $B$ is able to be accepted by a verifier with a probability upper than $\frac{k+i}{2k}$, $B$ can retrieve the secret key of the protocol from the public one with a probability greater than, $1-\frac{2^{n-k}-i}{(2^{n-k}+n-1)^i}\binom{n}{w}^i$, or find a collision for the hash function in polynomial time. 

\end{theo}

{\bf Sketch of proof :}\\

     Suppose a malicious prover M is able to answer $k+i$ challenges. By the pigeonhole principle he is able to answer $2i$ challenges of the form $\{(r_j,b), 1\leq j \leq i$ and $b \in \{0,1\}\}$.
Rewriting the commitment $c_3$ in two differents ways shows that he is able to construct a (i+1)-uplet $(c,z_1,\dots,z_i)$ solution of the following problem :

\begin{equation}
        \label{equa}
s_{r_j}=c+H\cdot z_j^t
\end{equation}

        with $wt(z_j)=w$, $s_{r_j}$ the syndrome of the public key $x$ shifted by $r_j$ positions, $c$ a constant vector and $1\leq j \leq i$. \\
        The next step consists in reducing the solutions of the problem (\ref{equa}) by increasing the value of the parameter $i$.
        We use probabilities to evaluate the size of the set of solutions and more particularly, the distribution of syndrome of words of weight $w$ for a double circulant code with adequate length, see \cite{GZ}. We deduce that a random tuple $(c,z_1,\dots,z_i)$ with $z_j$ a word of fixed weight $w$ for $1\leq j \leq i$ satisfies the set of equations (\ref{equa}) with probability equal to $\frac{1}{2^{n-k}+n-1}$. A careful probability analysis gives the bound described in the theorem. This probability depends on $i$ which is the number of conditions.  \\
        Notice that the tuple $(0,z_1,\dots,z_i)$ is a solution of the equation \ref{equa} with $z_j$ equal to the secret key shifted by block for $1 \leq j \leq i$. Since we choose $i$ such that the shifted secret key is the unique solution with a very strong probability, therefore a malicious
prover who knows how to answer in $k+i$ cases under $2k$ will be able to retrieve the secret key with a shift by block with a very strong probability (in practice the probability is chosen up to
$1-2^{-80}$).

\subsection{Zero-Knowledge}

This part of the proof consists in proving that no information can be deduce in polynomial time from an execution of the protocol more than the knowledge of the public data.
The idea is to prove that anyone can build a simulator of the protocol in polynomial time such that the result of the simulator cannot be distinguished from a real execution. \\
The simulator is build by anticipation by the challenges, for each round it is possible to make a valid instance by anticipation of the challenge $b$ only. This implies a construction
in twice the number of rounds of the protocol. \\
The case $b=0$ can be anticipated by the choice of $\sigma ^{'}$ a random permutation, $v$ a random word, $h_1=hash(\sigma ^{'})$ and $h_3=hash(\sigma ^{'}(vG+x_r))$. We notice that ($v$,$\sigma ^{'}$) and ($u+m_r$,$\sigma$) are indistinguishable.
The case $b=1$ can be anticipated by the choice of $v$ and $z$ such as $z$ is a word of weight $w$, $v=\pi(uG)$ with $\pi$ a random permutation, $u$ a random word, $h_2=hash(v)$ and
$h_3=hash(v+z)$. We notice that ($v$,$z$) and ($\sigma (uG)$,$\sigma (e_r)$) are indistinguishable.\\
The construction's cost of the simulator is negligible and does not affect the security parameters. When we use the commitment compression improvement the proof is different because of the complexity cost of
anticipation, in this case the construction's cost of the simulator is not negligible and it is more interesting to produce this improvement several times instead of one to not affect the security too much.

\subsection{Practical security of double circulant codes}

At the difference of the original Stern's scheme, our protocol is based
on decoding a random double-circulant matrix (SD-DC problem say), this problem 
at the difference of the SD problem, is not proven NP-hard 
(although a result is known on the hardness of decoding general quasi-cyclic 
codes).  Meanwhile in our case the problem appears to be hard since : 1) it has been proven
in \cite{GZ} that random double circulant codes rely on the GV bound, 2) it is not known, even with very structured
codes, how to decode a code up to the GV bound in polynomial time and at last, 3),
in practice, there is no known specialized algorithm which can do significantly better
(besides a small linear factor $n$) for solving the SD-DC problem. The situation is the same than
for lattices and ideal lattices compared to random lattices. In practice the best known 
algorithm to attack the SD-DC problem are the same than those for the SD problem (\cite{FS09}).

\section{Parameters for authentication and signature}

According to the security constraints for zero-knowledge discuss earlier  
	we choose as parameters $n=698,k=349,i=19,w=70$ for a security in $2^{81}$  
and a probability of cheating in $2^{-16}$.

\begin{table}[!h]

\centering

\caption{''Comparison between ZK scheme for a $2^{-16}$ cheating probability''}

\begin{tabular}{|c|c|c|}
	\hline
       & Stern 3 & Stern 5 \\
	\hline
   Rounds & 28 & 16 \\
	\hline
   Matrix size (bits)& 122500& 122500 \\
	\hline
   Public Id (bits)& 350 & 2450 \\
	\hline
   Secret key (bits)& 700 & 4900 \\
	\hline
   Communication (bits)& 42019 & 62272 \\
	\hline
	Prover's Computation & $2^{22.7}$op. in$\mathbb{F}_2$ & $2^{21.92}$op. in$\mathbb{F}_2$  \\
\hline
\end{tabular}

\end{table}

\begin{table}[!h]

\centering

%\caption{''Comparison between ZK scheme for a $2^{-16}$ cheating probability''}

\begin{tabular}{|c|c|c|}

	\hline

        Veron & CVE & New protocol  \\

	\hline

    28 & 16 & 18\\

	\hline

    122500 & 32768 & 350 \\

	\hline

    700 & 512 & 700 \\

	\hline

    1050 & 1024 & 700 \\

	\hline

    35486 & 31888 & 20080 \\

	\hline

 $2^{22.7}$op. in$\mathbb{F}_2$ & $2^{16}$mult. in$\mathbb{F}_{256}$ &  $2^{21}$op. in$\mathbb{F}_2$ \\

\hline

\end{tabular}

\end{table}

	For a security in $2^{100}$ we choose $n=838,k=419,i=20,w=86$ and for a security in $2^{128}$ we have $n=1094,k=547,i=14,w=109$.

$\bullet$ For signature, and a probability of cheating in $2^{80}$ it is sufficient to multiply by $5$
the previous data. Overall our double-circulant scheme permits to obtain a signature of length
93kb.

{\bf Remark}: it is possible to decrease even more the communication cost by using a constant weight encoding when sending $\sigma(e_r)$, the cost is then $k$ bits rather than $2k$ bits,
overall it decreases the authentication to 17kb and the signature to 79kb, but the encoding
comes with a complexity price.

\section{Conclusion}

In this paper we propose a new variation on Stern's authentication scheme. Our protocol   
permits to obtain a gain of more than 40\% compared to previous schemes and it is the first code based zero knowledge scheme
to obtain a signature length of less than 100kb with strong security 
and small size of keys.

\bibliography{biblio}

\bibliographystyle{alpha}

\end{document}